\documentstyle[12pt,fleqn]{article}
\gdef\journal#1,#2,#3,#4.{{\it #1~}{\bf #2} (#4) #3}

\def\be{\begin{equation}}\def\bea{\begin{eqnarray}}\def\beaa{\begin{eqnarray*}}
  \def\ee{\end{equation}}  \def\eea{\end{eqnarray}}  \def\eeaa{\end{eqnarray*}}

\def\la{\mathrel{\mathpalette\fun <}}

\def\fun#1#2{\lower3.6pt\vbox{\baselineskip0pt\lineskip.9pt
        \ialign{$\mathsurround=0pt#1\hfill##\hfil$\crcr#2\crcr\sim\crcr}}}
\begin{document}
\def\half{{\textstyle{ 1\over 2}}}
\def\frac#1#2{{\textstyle{#1\over #2}}}
\def\gsim{\mathrel{\raise.3ex\hbox{$>$\kern-.75em\lower1ex\hbox{$\sim$}}}}
\def\lsim{\mathrel{\raise.3ex\hbox{$<$\kern-.75em\lower1ex\hbox{$\sim$}}}}
\def\la{\bigl\langle} \def\ra{\bigr\rangle}
\def\cd{\!\cdot\!}
\def\a{\hat a}      \def\b{\hat b}      \def\c{\hat c}
\def\ab{\a\cd\b}    \def\ac{\a\cd\c}    \def\bc{\b\cd\c}
\def\cg{\cos\gamma} \def\ca{\cos\alpha} \def\cb{\cos\beta}
\def\gg{\hat\gamma}    \def\go{ \hat\gamma_1}
\def\gt{\hat\gamma_2}  \def\gth{\hat\gamma_3}
\def\got{ \hat\gamma_1\cd\hat\gamma_2} \def\ggo{ \hat\gamma\cd\hat\gamma_1}
\def\goth{\hat\gamma_1\cd\hat\gamma_3} \def\ggt{ \hat\gamma\cd\hat\gamma_2}
\def\gtth{\hat\gamma_2\cd\hat\gamma_3} \def\ggth{\hat\gamma\cd\hat\gamma_3}

\def\n{\hat n}       \def\no{\hat n_1}   \def\nt{\hat n_2}  \def\nth{\hat n_3}
\def\nont{\no\cd\nt} \def\nonth{\no\cd\nth} \def\ntnth{\nt\cd\nth}

\def\nogo{\no\cd\hat\gamma_1} \def\nogt{\no\cd\hat\gamma_2}
\def\nogth{\no\cd\hat\gamma_3}
\def\ntgo{\nt\cd\hat\gamma_1} \def\ntgt{\nt\cd\hat\gamma_2}
\def\ntgth{\nt\cd\hat\gamma_3}
\def\nthgo{\nth\cd\hat\gamma_1} \def\nthgt{\nth\cd\hat\gamma_2}
\def\nthgth{\nth\cd\hat\gamma_3}

\def\D{ {\Delta T \over T} }   \def\dO{d\Omega}

\begin{flushright}
SISSA REF. 193/93/A \\
DFPD 93/A/80 \\
December 1993 \\
\end{flushright}

\vspace{.2in}

\begin{center}
{\Large {\bf THE THREE--POINT CORRELATION FUNCTION}}

\vspace{.2in}

{\Large {\bf OF THE COSMIC MICROWAVE BACKGROUND}}

\vspace{.2in}

{\Large {\bf IN INFLATIONARY MODELS}}

\vspace{.3in}

{\bf Alejandro Gangui}$^1$, {\bf Francesco Lucchin}$^2$, {\bf Sabino
Matarrese}$^3$ and {\bf Silvia Mollerach}$^4$ \\

\vspace{.2in}

$^1${\em SISSA -- International School for Advanced Studies, \\
via Beirut 2 -- 4, 34013 Trieste, Italy}\\

$^2${\em Dipartimento di Astronomia Universit\`a di Padova, \\
vicolo dell'Osservatorio 5, 35122 Padova, Italy}\\

$^3${\em Dipartimento di Fisica ``Galileo Galilei" Universit\`a di Padova, \\
via Marzolo 8, 35131 Padova, Italy}\\

$^4${\em Theory Division CERN, CH--1211, Gen\`eve 23, Switzerland}\\

\end{center}

\vspace{.2in}

\begin{center}
Submitted to {\em The Astrophysical Journal}
\end{center}

\vspace{.1in}

\begin{center}
\section*{Abstract}
\end{center}

We analyze the temperature three--point correlation function and the
skewness of the Cosmic Microwave Background (CMB),
providing general relations in terms of multipole coefficients.
We then focus on applications to large angular scale anisotropies,
such as those measured by the {\em COBE} DMR, calculating
the contribution to these quantities from primordial, inflation generated,
scalar perturbations, via the Sachs--Wolfe effect. Using the techniques of
stochastic inflation we are able to provide a {\it universal} expression for
the ensemble averaged three--point function and for the corresponding
skewness, which accounts for all primordial second--order effects. These
general expressions would moreover apply to any situation where
the bispectrum of the primordial gravitational potential has a {\em
hierarchical} form. Our results are then specialized to a number of relevant
models: power--law inflation driven by an exponential potential, chaotic
inflation with a quartic and quadratic potential and a particular case
of hybrid inflation. In all these cases non--Gaussian effects are small: as
an example, the {\em mean} skewness is much smaller than the
cosmic {\em rms} skewness implied by a Gaussian temperature fluctuation field.

\newpage

\section{Introduction}

The recent detection of large--scale Cosmic Microwave Background
(CMB) anisotropies by {\em COBE} (Smoot et al. 1992) has provided us
with a unique opportunity to probe the properties of the primordial
perturbations which affect the CMB temperature distribution through the
Sachs--Wolfe effect (Sachs \& Wolfe 1967). Future years of observation will
reduce the level of the noise, thus making possible to extract more and more
statistical information from the data. Besides the classical analysis
in terms of the {\em rms} temperature fluctuation on the smoothing scale of
the DMR instrument, the angular two--point correlation function and the lowest
order multipoles (Smoot et al. 1992; Wright et al. 1992), preliminary results
have already appeared on a measurement of the three--point function and
skewness on the {\em COBE} map (Hinshaw et al. 1993; Smoot et al. 1993).
The three--point correlation function
of CMB fluctuations, and its limit at zero angular separation, the skewness,
provide an estimate of the size of possible deviations from a Gaussian
behaviour of the primordial perturbation field.

There are at present two main theories on the origin of primordial
perturbations, namely that they originated from quantum fluctuations
of scalar fields during an early inflationary era or from topological
defects produced by a phase transition in the early universe. Present
observational data have not yet been able to rule out one of them. It
was generally thought that the main difference between the two models is
the statistical distribution of the resulting fluctuations:
inflationary models were claimed to predict a Gaussian distribution,
while topological defects a non--Gaussian one. However, it is now clear
that, at least for the cosmic string case, even if the effect of a
single string is clearly non--Gaussian, the combined effect of a large
number of them results in a {\em quasi--Gaussian} distribution. On the other
hand, it has been realized that inflationary models also predict
small deviations from Gaussianity. Hence, it is necessary to make more
quantitative estimates of the non--Gaussian features expected in each
model if one intends to use the predicted statistical distribution of
CMB anisotropies to discriminate between them.

Besides instrumental noise and contamination
by Galactic emission, any analysis of large--scale anisotropies is however
limited by one more source of indetermination:
the so--called ``cosmic variance" (Abbott \& Wise 1984; Smoot et al. 1992),
namely the impossibility of making observations in more than one universe,
which severely limits our ability to extract intrinsic non--Gaussian signals
from the data, especially for anisotropy patterns dominated by low--order
multipoles (Scaramella \& Vittorio 1991). This is particularly important when
considering the temperature
three--point correlation function of the {\em COBE} map, since most present
theoretical models of the primordial density fluctuation process predict
rather small non--Gaussian effects on such large scales, i.e. a
quasi--Gaussian anisotropy pattern.
Actually, the large beam--width of the {\em COBE} DMR experiment blurs any
information on smaller scales, which makes it practically insensitive to the
type of non--Gaussian signatures produced by topological defects.
Cosmic strings predict relevant non--Gaussian features in the CMB
at angular scales below a few arcminutes (Moessner, Perivolaropoulos
\& Brandenberger 1993). This makes even more interesting obtaining a
quantitative prediction on the non--Gaussian CMB features which are expected
to be produced on very large angular scales in inflationary models. In fact,
every sensible inflation model will produce some small but
non--negligible non--Gaussian effects, both
by the self--interaction of the {\em inflaton} field,
and by the local back--reaction of its
self--gravity. It is our interest here to quantify this prediction,
with the largest possible generality, so that observational results
on the CMB angular three--point function can be used as a further test
on the nature of the primordial perturbation process.

Unfortunately, we will find that single--field inflation models
generally imply mean values for the skewness which are well below
the cosmic {\em rms} skewness of a Gaussian field, which confirms and
generalizes earlier results based on a simple toy--model (Falk, Rangarajan \&
Srednicki 1993; Srednicki 1993).
In this sense, looking at anisotropies on smaller angular scales than
{\em COBE} would probably appear a more promising strategy.
Coulson, Pen \& Turok (1993) and Coulson et al. (1993) consider
degree--scale anisotropies produced by non--Gaussian field--ordering
theories of structure formation. The results of the present paper would also
apply to these intermediate scales, only provided the window function
appropriate to the specific experiment is used.

Falk et al. (1993) first gave a quantitative estimate of the size of
non--Gaussian effects through a calculation of the three--point CMB
correlation function from perturbations generated in a simple model, where the
inflaton has cubic self--interactions. Luo \& Schramm (1993) argued that
secondary anisotropies could also play a non--negligible role in this respect.

Our paper is the first one where the problem is considered in a totally
general and self--consistent way. We use the stochastic approach
to inflation (e.g. Starobinskii 1986; Goncharov, Linde \& Mukhanov 1987), as
a technical tool to self--consistently account for all second--order effects
in the generation of scalar field fluctuations during inflation and their
giving rise to curvature perturbations. We also properly account for the
non--linear relation between the inflaton fluctuation and the peculiar
gravitational potential. Our derivation moreover removes a non--realistic
restriction to purely de Sitter background made by Falk et al. (1993), which is
especially important when non--Zel'dovich perturbation spectra are considered.

The plan of the paper is as follows. In Section 2 we give completely
general definitions for the CMB three--point function in a single
sky in terms of multipole coefficients. We stress that these results actually
apply to whichever type of temperature fluctuations, as well
as to any angular scale. We also give what we believe is a more operational
definition of angular three--point function than has been considered so far
(e.g. Luo 1993a,b). In the last part of this section we anticipate the results
obtained from the Sachs--Wolfe effect due to the quasi--Gaussian
perturbations of the gravitational potential generated during inflation
and we discuss the effects of the cosmic variance on the possibility of
observing a non--zero mean value for the skewness. As a by--product of this
analysis we compute the ``dimensionless" {\em rms} skewness for a large range
of the primordial spectral index of density fluctuations $n$.
All the results reported at this level only depend on three quantities:
i) the {\em rms} quadrupole amplitude ${\cal Q}$, which is fixed
by observational data; ii) a model--dependent parameter $\Phi_3 \sim$ {\em a
few}, related to the amplitude of the skewness;
iii) the rotationally invariant multipole coefficients ${\cal C}_\ell$
(suitably normalized to the quadrupole, $\ell=2$), whose specific dependence
on $\ell$ is fixed by the spectral index $n$. Note
that inflation is able to produce perturbation spectra with essentially all
values of $n$ around unity (e.g. Mollerach, Matarrese \& Lucchin 1993), as
it can be needed to match the {\em COBE} data (Smoot et al. 1992) with
observations on smaller scales.
The technical derivation in the frame of stochastic inflation of the
connected
two-- and three--point function (or its Fourier counterpart, the bispectrum),
for the inflaton field first and the local gravitational potential next, is
reported in Section 3.
Section 4 deals with the zero--lag limit of the three--point function:
the skewness, for which we provide a universal inflationary expression,
which we then specialize to some of the most popular inflaton potentials:
exponential (Lucchin \& Matarrese (1985)), quartic and quadratic
(Linde 1983, 1985) as well as a simple potential for hybrid inflation (Linde
1993). Section 5 contains some general conclusions.

\section{The CMB three--point correlation function}

Our aim here is to compute the CMB temperature three--point
correlation function implied by possible departures from a Gaussian
behaviour of the primordial peculiar gravitational potential.
As a first step we will define the connected two-- and three--point
correlation functions of the CMB temperature as measured by a given observer,
i.e. on a single microwave sky. Let us then define the temperature fluctuation
field $\D(\vec x ;\gg) \equiv (T(\vec x ;\gg) - T_0(\vec x))/T_0(\vec x)$,
where $\vec x$ specifies the position of the observer, the unit vector
$\gg$ points in a given direction from $\vec x$ and
$T_0(\vec x) \equiv \int {\dO_{\gg}\over 4\pi} T(\vec x ;\gg)$
represents the mean temperature of the CMB measured by that observer
(i.e. the {\em monopole} term).

The angular two--point correlation function $C_2(\vec x; \alpha)$ measured
by an observer placed in $\vec x$ is the average product of
temperature fluctuations in two directions $\gg_1$ and $\gg_2$ whose angular
separation is $\alpha$; this can be written as
\be
\label{2cor}
C_2 (\vec x; \alpha) =
\int {\dO_{\gg_1}\over 4\pi} \int {\dO_{\gg_2}\over 2\pi}
\delta\bigl(\gg_1 \cdot \gg_2 - \cos \alpha\bigr)
\D(\vec x ;\gg_1) \D(\vec x ;\gg_2) \ .
\ee
As well known, in the limit $\alpha \to 0$ one recovers the CMB variance
$C_2 (\vec x ) = \int {\dO_{\gg}\over 4\pi} [\D(\vec x ;\gg)]^2$.
Expanding the temperature fluctuation in spherical harmonics
\be
\D(\vec x ;\gg) = \sum_{\ell=1}^\infty \sum_{m=-\ell}^\ell a_\ell^m (\vec x)
{\cal W}_\ell Y_\ell^m(\gg),
\ee
and writing the Dirac delta function as a
completeness relation for Legendre polynomials $P_\ell$, we easily arrive
at the expression
\be
\label{2cor'}
C_2 (\vec x; \alpha) =
{1 \over 4\pi} \sum_{\ell} P_\ell (\cos \alpha) Q_\ell^2 (\vec x)
{\cal W}^2_\ell \ ,
\ee
where $Q_\ell^2 = \sum_{m=-\ell}^\ell \vert a_\ell^m \vert^2$.
In the previous expressions ${\cal W}_\ell$ represents the window function
of the specific experiment. Setting ${\cal W}_0 = {\cal W}_1=0$ automatically
accounts for both monopole and dipole subtraction; for $\ell \geq 2$ one can
take ${\cal W}_{\ell}\simeq\exp\left[-\frac12 \ell({\ell}+1)\sigma^2\right]$,
where $\sigma$ is the dispersion of the antenna--beam profile, which
measures the angular response of the detector (e.g. Wright et al. 1992).
In some cases the quadrupole term is also subtracted from the maps (e.g. Smoot
et al. 1992); in this case we also set ${\cal W}_2=0$.

The analogous expression for the angular three--point correlation function
is obtained by taking the average product of
temperature fluctuations in three directions $\gg_1$, $\gg_2$ and $\gg_3$
with fixed angular separations $\alpha$ (between $\gg_1$ and $\gg_2$), $\beta$
(between $\gg_2$ and $\gg_3$) and $\gamma$ (between $\gg_1$ and $\gg_3$);
these angles have to satisfy the obvious inequalities $\vert \alpha - \gamma
\vert \leq \beta \leq \alpha + \gamma$.
One then has
\bea
\label{3cor}
\lefteqn{
C_3 (\vec x; \alpha,\beta,\gamma) =
\int {\dO_{\gg_1}\over 4\pi} \int_0^{2\pi} {d \varphi_{12} \over 2 \pi}
\int_{-1}^1 d\cos \vartheta_{12}
\delta\bigl(\cos \vartheta_{12} - \cos \alpha\bigr)
\int_{-1}^1 d\cos \vartheta_{23}
}
\nonumber \\
&  &
\times \delta\bigl(\cos \vartheta_{23} - \cos \beta\bigr)
\!\int_{-1}^1 \!\! d\cos \vartheta_{13}
\delta\bigl(\cos \vartheta_{13} - \cos \gamma\bigr)
\D(\vec x ;\gg_1) \D(\vec x ;\gg_2) \D(\vec x ;\gg_3),
\eea
where $\cos \vartheta_{\alpha\beta} \equiv \gg_\alpha \cdot \gg_\beta$
and $\varphi_{12}$ is the azimuthal angle of $\gg_2$ on the plane
orthogonal to $\gg_1$. The above relation can be rewritten in a form
analogous to Eq.(\ref{2cor}), namely
\bea
\label{3cor'}
\lefteqn{
C_3 (\vec x; \alpha,\beta,\gamma) = N(\alpha,\beta,\gamma)
\int {\dO_{\gg_1}\over 4\pi} \int {\dO_{\gg_2}\over 2\pi} \int
{\dO_{\gg_3}\over 2} \delta\bigl(\gg_1 \cdot \gg_2 - \cos \alpha\bigr)
}
\nonumber \\
&  &
\times
\delta\bigl(\gg_2 \cdot \gg_3 - \cos \beta\bigr)
\delta\bigl(\gg_1 \cdot \gg_3 - \cos \gamma\bigr)
\D(\vec x ;\gg_1) \D(\vec x ;\gg_2) \D(\vec x ;\gg_3) \ ,
\eea
where $N(\alpha,\beta,\gamma) \equiv \sqrt{1 - \!\cos^2\!\alpha -
\!\cos^2\!\beta - \!\cos^2\!\gamma + \!2
\!\cos\!\alpha\!\cos\!\beta\!\cos\!\gamma}$.\footnote{
To show that the latter expression is correctly normalized one can
expand the delta functions in Legendre polynomials and use
the relation $\sum_\ell (2 \ell +1) P_\ell(x) P_\ell(y) P_\ell(z)
= {2 \over \pi} (1-x^2-y^2-z^2+2xyz)^{-1/2}$.}
Setting $\alpha=\beta=\gamma=0$ in these general expressions one obtains the
CMB {\em skewness} $C_3 (\vec x ) = \int {\dO_{\gg}\over 4\pi}
[\D(\vec x ;\gg)]^3$. Also useful are the {\em equilateral} three--point
correlation function (e.g. Falk et al. 1993) and the
{\em collapsed} one (e.g. Hinshaw et al. 1993 and references therein),
corresponding to the choices
$\alpha=\beta=\gamma$, and $\alpha=\gamma$, $\beta=0$, respectively.
Alternative statistical estimators, more suited to discriminate bumpy
non--Gaussian signatures in noisy data, have been recently introduced by
Graham et al. (1993).
In all the above formulas, full--sky coverage was assumed, for simplicity.
The effects of partial sky coverage on some of the statistical quantities
considered here are discussed in detail by Scott, Srednicki \& White (1993).

Following the procedure used above for $C_2(\vec x;\alpha)$, we
can rewrite the three--point function in the form
\bea
\label{3cor''}
\lefteqn{
C_3 (\vec x; \alpha,\beta,\gamma) = N(\alpha,\beta,\gamma) {\pi \over 2}
\sum_{\ell_1,\ell_2,\ell_3}\sum_{m_1,m_2,m_3}
a_{\ell_1}^{m_1} a_{\ell_2}^{m_2}
{a_{\ell_3}^{m_3}}^* {\cal W}_{\ell_1} {\cal W}_{\ell_2} {\cal W}_{\ell_3}
}
\nonumber \\
&  &
\times
\sum_{j,k,\ell} \sum_{m_j,m_k,m_\ell} P_j(\cos\alpha) P_k(\cos\beta)
P_\ell(\cos\gamma) {\cal H}_{j\ell\ell_1}^{m_j m_\ell m_1}
{\cal H}_{kj\ell_2}^{m_k m_j m_2}
{\cal H}_{k\ell\ell_3}^{m_k m_\ell m_3} \ ,
\eea
where the coefficients ${\cal H}_{\ell_1\ell_2\ell_3}^{m_1 m_2 m_3}
\equiv \int \dO_{\gg} {Y_{\ell_1}^{m_1}}^*(\gg) Y_{\ell_2}^{m_1}(\gg)
Y_{\ell_3}^{m_3}(\gg)$, which can be easily expressed in terms of
Clebsch--Gordan coefficients (e.g. Messiah 1976), are only non--zero if
the indices $\ell_i$, $m_i$ ($i=1,2,3,$) fulfill the relations:
$\vert \ell_j - \ell_k \vert \leq  \ell_i \leq \vert \ell_j + \ell_k \vert$,
$\ell_1 + \ell_2 + \ell_3 = even$ and $m_1 = m_2 + m_3$.
The collapsed three--point function measured by the
observer in $\vec x$ reads
\be
\label{skew}
C_3 (\vec x;\alpha) = {1 \over 4 \pi}
\sum_{\ell_1,\ell_2,\ell_3}\sum_{m_1,m_2,m_3}
P_{\ell_1}(\cos\alpha)
a_{\ell_1}^{m_1} a_{\ell_2}^{m_2}
{a_{\ell_3}^{m_3}}^* {\cal W}_{\ell_1} {\cal W}_{\ell_2} {\cal W}_{\ell_3}
{\cal H}_{\ell_3\ell_2\ell_1}^{m_3 m_2 m_1}
\ee
which, for $\alpha=0$, gives a useful expression for the skewness.

So far our expressions have been kept completely general, they would apply
to whatever source of temperature fluctuations in the sky, through
suitable (usually statistical) relations for the product of three multipole
coefficients $a_\ell^m$ appearing in Eqs.(\ref{3cor''}) and (\ref{skew}),
and to whatever angular scale, through the specific choice of window functions
${\cal W}_\ell$.
However, in what follows we shall only deal with large angular scale
anisotropies originated from primary perturbations in the gravitational
potential $\Phi$ on the last scattering surface via the Sachs--Wolfe effect
(Sachs \& Wolfe 1967). In that case
$\D(\vec x ;\gg) = {1 \over 3} \Phi( \vec x + r_0 \gg)$, where $r_0=2/H_0$
is the horizon distance and $H_0$ the Hubble constant. As noticed by Luo
\& Schramm (1993), secondary anisotropies produced during the mildly
non--linear evolution of perturbations through the Rees--Sciama effect
(Rees \& Sciama 1968; Mart{\'\i}nez--Gonz\'alez, Sanz \& Silk 1992), also
imply a non--zero contribution to the connected three--point function, which is
still there when the underlying primordial linear gravitational potential is
Gaussian. Furthermore, these authors suggest that this contribution should
strongly dominate over the primary inflationary one.
However, a more detailed analysis (Mollerach et al., in preparation) shows
that this is not the case on the angular scales probed by the {\em COBE} DMR.
Therefore, in what follows we will only refer to the primary contribution
to the three--point function. A recent analysis of the three--point function
and the skewness of the {\em COBE} data has been performed by Hinshaw et al.
(1993) and Smoot et al. (1993), who find no statistical evidence for
non--Gaussian signatures beyond those implied by the cosmic variance:
while consistent with the random--phase hypothesis, these data can only
rule out strongly non--Gaussian fluctuations on very large scales.
On smaller scales, Graham et al. (1993), analyzing the {\em UCSB} SP91
experiment, were able to detect non--Gaussian features, which,
however, might also be of non--cosmological origin.

To obtain definite predictions for the statistics described above,
one needs to exploit the random nature of the multipole coefficients
$a_\ell^m$. In our case, these coefficients should be considered as
zero--mean non--Gaussian random variables whose statistics should in
principle be deduced from that of the gravitational potential. In such a case
one should either obtain, by Monte--Carlo simulations, different realization of
the sky corresponding to different ``cosmic observers",
or analytically compute theoretical ensemble expectation values.
The latter procedure will be followed here to obtain the mean two-- and
three--point functions. These expectation values are of course observer-- i.e.
$\vec x$--independent and can only depend upon the needed number of
angular separations.

In the frame of the inflationary model, the calculations reported
in the following section lead to general expressions for the
mean two-- and three--point functions of the primordial gravitational
potential, namely
\be
\label{newer}
\la \Phi(r_0 \go) \Phi(r_0 \gt) \ra = {9 \pi {\cal Q}^2 \over 5 (2\pi)^2}
\sum_{\ell\ge 0} (2 \ell+1) P_\ell(\gg_1\cdot\gg_2) {\cal C}_\ell
\ee
and
\bea
\label{eq717}
\lefteqn{
\la \Phi(r_0 \go) \Phi(r_0 \gt) \Phi(r_0 \gth) \ra
= {81\pi^2 {\cal Q}^4 \over 25 (2\pi)^4} \Phi_3
\sum_{j, \ell \ge 0} (2j+1)(2\ell+1) {\cal C}_j {\cal C}_\ell
	}
\nonumber \\
&  &
\times
\bigl[ P_j(\go \cdot \gth) P_\ell(\go \cdot \gt) +
       P_j(\gt \cdot \go) P_\ell(\gt \cdot \gth) +
       P_j(\gth \cdot \go) P_\ell(\gth \cdot \gt) \bigr] \ ,
\eea
where $\Phi_3$ is a model--dependent coefficient.
The $\ell$--dependent coefficients ${\cal C}_\ell$ are
defined by $\la Q_\ell^2 \ra \equiv {(2 \ell + 1)  \over 5} {\cal Q}^2
{\cal C}_\ell$, with ${\cal Q} = \la Q_2^2 \ra^{1/2}$ the {\em rms}
quadrupole, and are
related to the gravitational potential power--spectrum $P_\Phi(k)$ through
${\cal C}_\ell = \int_0^\infty dk k^2 P_\Phi(k) j_\ell^2(k r_0) /
\int_0^\infty dk k^2 P_\Phi(k) j_2^2(k r_0)$,
where $j_\ell$ is the $\ell$--th order spherical Bessel
function.\footnote{The possible infrared
divergence of this expression for $\ell=0$ has no practical effect on
observable quantities, since the monopole is always removed.}
The {\em rms} quadrupole is simply related to the
quantity
$Q_{rms-PS}$ defined by Smoot et al. (1992): ${\cal Q} =\sqrt{4\pi}
Q_{rms-PS}/T_0$. For the scales of interest we can make the
approximation $P_\Phi(k) \propto k^{n-4}$, where $n$ corresponds to the
primordial index of density fluctuations (e.g. $n=1$ is the Zel'dovich,
scale--invariant case), in which case (e.g. Bond \& Efstathiou 1987;
Fabbri, Lucchin \& Matarrese 1987)
\be
\label{cl}
{\cal C}_\ell
=
{\Gamma (\ell+\frac{n}{2}-\frac{1}{2})
\Gamma\left(\frac{9}{2}-\frac{n}{2}\right)
\over \Gamma\left(\ell+\frac{5}{2}-\frac{n}{2}\right)
\Gamma (\frac{3}{2}+\frac{n}{2})}
\ee

The equations above allow to compute the angular spectrum,
\be
\la a_{\ell_1}^{m_1} {a_{\ell_2}^{m_2}}^* \ra =
\delta_{\ell_1\ell_2} \delta_{m_1 m_2} { {\cal Q}^2 \over 5} {\cal C}_{\ell_1}
\ ,
\ee
and the angular bispectrum,
\be
\la a_{\ell_1}^{m_1} a_{\ell_2}^{m_2} {a_{\ell_3}^{m_3}}^* \ra
= {3 {\cal Q}^4 \over 25} \Phi_3 \bigl[ {\cal C}_{\ell_1} {\cal C}_{\ell_2} +
{\cal C}_{\ell_2} {\cal C}_{\ell_3} +
{\cal C}_{\ell_3} {\cal C}_{\ell_1}
\bigr] {\cal H}_{\ell_3 \ell_1 \ell_2}^{m_3 m_1 m_2} \ .
\ee
Replacing the latter expression into Eq.(\ref{3cor''}) we obtain the general
form of the mean three--point correlation function. Some simplifications occur
for the collapsed three--point function, for which we obtain
\bea
\label{eq57}
\lefteqn{
\la C_3 (\alpha) \ra = {3 {\cal Q}^4 \over 25 (4\pi)^2} \Phi_3
\sum_{\ell_1,\ell_2,\ell_3}\!\! (2 \ell_1 +1) (2 \ell_2 +1) (2 \ell_3 +1)
P_{\ell_1}(\cos\alpha) \bigl[ {\cal C}_{\ell_1} {\cal C}_{\ell_2} +
{\cal C}_{\ell_2} {\cal C}_{\ell_3} +
{\cal C}_{\ell_3} {\cal C}_{\ell_1} \bigr]
	}
\nonumber \\
&  &
\times
{\cal W}_{\ell_1} {\cal W}_{\ell_2} {\cal W}_{\ell_3}
{\cal F}_{\ell_1 \ell_2 \ell_3}
\eea
where the coefficients ${\cal F}_{\ell_1 \ell_2 \ell_3} \equiv
(4\pi)^{-2}\int\dO_{\gg} \int\dO_{\gg'} P_{\ell_1}(\gg\cdot\gg') P_{\ell_2}
(\gg\cdot\gg') P_{\ell_3}(\gg\cdot\gg')$ may be suitably expressed in terms of
products of factorials of $\ell_1$, $\ell_2$ and $\ell_3$, using standard
relations for Clebsch--Gordan coefficients, by noting that
${\cal F}_{k \ell m} = \left(^{k~\ell~m}_{0~0~0}\right)^2$ (cf. Messiah
1976). The CMB mean skewness $\la C_3(0) \ra$ immediately follows from the
above equation for $\alpha=0$. A plot of the angular dependence of the
collapsed three--point function above, normalized to the skewness, is reported
in Figure 1, for typical values of the spectral index $n$.

It was first realized by Scaramella \& Vittorio (1991) that detecting a
non--zero three--point function or skewness for temperature
fluctuations in the sky cannot be directly interpreted as a signal for
intrinsically non--Gaussian perturbations. In fact, even a Gaussian
perturbation field has non--zero chance to produce a non--Gaussian sky
pattern. This problem is related to what is presently known as cosmic
variance, and is particularly relevant for fluctuations on large angular
scales, i.e. for low--order multipoles of the temperature fluctuation field.
One way to quantify this effect is through the {\em rms} skewness
of a Gaussian field  $\la C_3^2(0) \ra _{Gauss}^{1/2}$.
It is easy to find
\be
\la C_3^2(0) \ra _{Gauss} = 3\int^1_{-1}d \cos\alpha \la C_2(\alpha) \ra ^3 \ .
\ee
Scaramella \& Vittorio (1991) and, more
recently, Srednicki (1993), focused on the most popular case
of a scale--invariant spectrum, $n=1$ and
the {\em COBE} DMR window function, corresponding to a Gaussian with
dispersion $\sigma=3^\circ\llap.2$ (e.g. Wright et al. 1992). We will be
interested here in the same quantity, but for various values of $n$.
In Figure 2 we have plotted the normalized {\it rms} skewness
$\la C_3^2(0) \ra_{Gauss}^{1/2} / \la C_2(0)\ra ^{3/2}$
as a function of the spectral index, both including and removing the
quadrupole: in both cases this ratio is in the range $0.1 - 0.3$ for
interesting values of $n$. The values obtained for $n=1$, both with and
without the $\ell=2$ contribution, are identical to those given by
Srednicki (1993), who adopted the same smoothing angle (the slightly different
definition of window function cannot affect our dimensionless skewness
ratios). As a rough criterion, we can
conclude that, in order to detect primordial non--Gaussian signatures,
$\la C_3(0) \ra$ must be at least of the same order as $\la C_3^2(0)
\ra_{Gauss}^{1/2}$.
On the other hand, as we will see below, single--scalar--field
inflationary models generally lead to skewness ratios
$\la C_3(0) \ra/ \la C_2(0) \ra^{3/2} \lsim  10^{-4}$, so that their
non--Gaussian features cannot be distinguished from the cosmic {\em rms}
skewness.
In this sense the quasi--Gaussian inflationary predictions for the CMB
anisotropies are in full agreement with the recent analysis of the
three--point function and the skewness from {\em COBE} data
(Hinshaw et al. 1993; Smoot et al. 1993).

\section{Stochastic inflation and the statistics of the gravitational
potential}

Now we will show the validity of Eq.(\ref{eq717})
for the three--point correlation function of the
gravitational potential due to perturbations produced during an
inflationary epoch in the early universe. This calculation will provide
the primordial power--spectrum, reflecting into the $\ell$ dependence of
the ${\cal C}_\ell$ coefficients, as well as a general expression for the
factors $\cal Q$ and $\Phi_3$ above.
In order to take into account all the effects contributing to a non--vanishing
primordial three--point correlation function of $\Phi$, we will perform
the computation in two steps. In Section 3.1 we compute the three--point
function for the inflaton field perturbation $\delta \phi$. The most
convenient way to perform this calculation is in the frame of the stochastic
approach to inflation (Starobinskii 1986, Goncharov et al. 1987),
which naturally takes into account all the multiplicative effects in the
inflaton dynamics that are responsible for the non--Gaussian features.
Then, in Section 3.2 we compute the extra--contribution to the three--point
function of the gravitational potential that arises due to the non--linear
relation between $\Phi$ and $\delta \phi$. This effect has been previously
noticed by Barrow \& Coles (1990) and Yi \& Vishniac (1993).

\subsection{The inflaton bispectrum}

To study the dynamics of the inflaton, we will apply the stochastic approach.
This is based on defining a coarse--grained inflaton field
$\phi(\vec x,\alpha)$, obtained by
suitable smoothing of the original quantum field over a scale larger
than the Hubble radius size, whose dynamics is described by
a multiplicative Langevin--type equation. This is obtained by adding
to the classical equation of motion a Gaussian noise term whose amplitude is
fixed by the {\em rms} fluctuation of the scalar field at Hubble radius
crossing,
\be
\label{lange}
{\partial \phi(\vec x, \alpha) \over \partial\alpha}=
- {V'(\phi) \over \kappa^2 V(\phi)} +
{H(\phi) \over 2\pi} \eta(\vec x, \alpha) \ ,
\ee
where $V(\phi)$ is the inflaton potential, primes denote differentiation
with respect to $\phi$ and $\kappa \equiv \sqrt{8\pi G} =
\sqrt{8\pi}/ m_P$, with $m_P$ the Planck mass.
The Hubble parameter here should be consistently calculated from the local
energy density of the coarse--grained inflaton.
The noise term $\eta$ has zero
mean and autocorrelation function (e.g. Mollerach et al. 1991)
\be
\label{auto}
\la
\eta (\vec x, \alpha)
\eta (\vec x', \alpha')
\ra
=
\hbar
j_0 ( q_s(\alpha) \vert \vec x - \vec x'\vert)
\delta (\alpha -\alpha') \ .
\ee
The use of the time variable $\alpha = \ln \left(a/a_*\right)$ in this
equation has been motivated by Starobinskii (1986), who noticed
that $\alpha$ accounts for the possible time dependence of the Hubble
parameter. This is particularly relevant when general,
i.e. non--de Sitter, inflation is studied. We also defined the coarse--grained
domain size through the comoving wave--number
$q_s(\alpha) \equiv \epsilon H(\alpha) a(\alpha)$, with $\epsilon$
a number smaller than one, $H(\alpha) \equiv H(\phi_{cl}(\alpha))$, with
$\phi_{cl}(\alpha)$ the homogeneous classical solution of the Langevin equation
(i.e., that obtained with the noise term ``switched" off). Finally the
scale--factor $a(\alpha)$ is obtained by integration of $H(\alpha)$.

The stochastic dynamics of the coarse--grained field
within a single coarse--graining domain (i.e., for $\vec x = \vec x'$)
can be studied in terms of the Fokker--Planck equation for the probability
distribution function of $\phi$.
In our case, instead, since we are interested also in spatial correlations
of the field, we will solve directly the Langevin equation above.
To the aim of computing the three--point function of $\phi$, a second--order
perturbative expansion around the classical solution
is enough. We will require that the potential $V(\phi)$ is a
smooth function of its argument, which translates into
requiring well defined values for the steepness of the potential
$X(\alpha) \equiv X(\phi_{cl}(\alpha)) \equiv m_P V'(\phi_{cl}) /
V(\phi_{cl})$
and its derivatives (Turner 1993) throughout the range of relevant scales.
Apart from this requirement we keep the analysis general; only at the end
we will apply our results to some specific inflationary potentials.
We first expand $V(\phi)$ around $\phi_{cl}$, up to second order in
$\delta\phi(\vec x,\alpha) \equiv \phi(\vec x,\alpha) - \phi_{cl}(\alpha)$,
(i.e. up to order $\hbar$), $V(\phi) = V(\phi_{cl}) + V'(\phi_{cl})
\delta\phi + \half V''(\phi_{cl}) \delta\phi^2 + \cdots$. Replacing this into
the Langevin equation we obtain
\be
\label{eq1}
{\partial \delta\phi(\vec x,\alpha) \over\partial\alpha} = A(\alpha)
\delta\phi(\vec x,\alpha) + B(\alpha) \delta\phi^2(\vec x,\alpha) +
\left[ D_1(\alpha) + D_2(\alpha) \delta\phi(\vec x,\alpha) \right]
\eta (\vec x,\alpha) \ ,
\ee
where we have used $\partial \phi_{cl}/ \partial\alpha = - V'(\phi_{cl}) /
\kappa^2 V(\phi_{cl})$. In Eq.(\ref{eq1}) we defined
\be
A = -{m_P\over 8\pi} X'
{}~ ; ~
B = -{m_P\over 16\pi} X''
{}~ ; ~
D_1 = {H(\alpha) \over 2\pi}
{}~ ; ~
D_2 = {H(\alpha)X \over 4 \pi m_P} \ .
\ee
Let us now split the field perturbation as $\delta\phi =
\delta\phi_1 + \delta\phi_2$ of ${\cal O}(\hbar^{1/2})$ and
${\cal O}(\hbar)$, respectively. We can also define a rescaled
variable $\tilde \eta \equiv (H(\alpha)/2\pi) \eta$.
We then find
\be
\label{eq94}
\delta\phi_1(\vec x ,\alpha) = X(\alpha)\int_0^\alpha
d\alpha' X^{-1}(\alpha') \tilde \eta (\vec x ,\alpha')
\ee
\be
\delta\phi_2(\vec x ,\alpha) = X(\alpha)
\int_0^{\alpha} d\alpha'
\left[
{B(\alpha')\over X(\alpha')} \delta\phi_1^2 (\vec x ,\alpha') +
{1\over 2 m_P}
\delta\phi_1 (\vec x ,\alpha') \tilde \eta (\vec x ,\alpha')
\right] \ .
\ee
Let us now calculate the connected three--point correlation function of
$\delta\phi$. The lowest order non--vanishing contribution
(${\cal O}(\hbar^2)$) reads
\bea
\label{eqale24}
\lefteqn{\la  \delta\phi (\vec x_1, \alpha_1)
\delta\phi (\vec x_2, \alpha_2)
\delta\phi (\vec x_3, \alpha_3) \ra =
}
\nonumber \\
&  &
\!X(\alpha_3) \!\int_0^{\alpha_3}\! d\alpha'
\left[{B(\alpha')\over X(\alpha')}
\la \delta\phi_1 (\vec x_1, \alpha_1) \delta\phi_1 (\vec x_3, \alpha')
\ra
\la \delta\phi_1 (\vec x_2, \alpha_2) \delta\phi_1 (\vec x_3, \alpha')
\ra\! +\! [\vec x_1\!\leftrightarrow\!\vec x_2 ] \right]
\nonumber \\
&  &
+ {X(\alpha_3)\over 2 m_P}\! \int_0^{\alpha_3}\! d\alpha'
\left[ \la \delta\phi_1 (\vec x_1, \alpha_1) \delta\phi_1 (\vec x_3, \alpha')
\ra \la \delta\phi_1 (\vec x_2, \alpha_2) \tilde \eta  (\vec x_3, \alpha')
\ra\! +\! [\vec x_1\!\leftrightarrow\!\vec x_2 ]
\right]
\nonumber \\
&  & + ~2 \times 2 ~terms \ .
\eea
The term proportional to $X(\alpha_3)/ 2 m_P$ in the r.h.s. of this
equation can be recast in the form
\be
\label{eq92}
{X(\alpha_2)X(\alpha_3)\over 2 m_P}\!
\int_0^{\alpha_{min}}\!\! d\alpha'
X^{-1}(\alpha')\!
{H^2(\alpha')\over (2\pi)^2}\!
\la
\delta\phi_1 (\vec x_1, \alpha_1)
\delta\phi_1 (\vec x_3, \alpha')
\ra
{}~j_0 (q_s(\alpha')\vert\vec x_2 -\vec x_3\vert)
\ee
where we defined $\alpha_{min}\equiv {\rm min}[\alpha_3,\alpha_2]$.
We need now to compute the $\delta\phi$ auto--correlation function.
To this aim, recalling that $\alpha$ is the time when the perturbation
wavelength
$\sim a/q$ equals the size of the coarse--graining domain,
we change the integration variable in Eq.(\ref{eq94})) from $\alpha$ to
$q = \sqrt{8\pi V(\alpha)/ 3}~ q_* e^{\alpha} / H_* m_P$.
The subscript $*$ denotes quantities evaluated at the time
when we start to solve the Langevin equation; this is chosen in such a way
that the patch of the universe where we live is homogeneous on a scale
slightly above our present horizon (see e.g. the discussion by Mollerach et
al. 1991). We then find
\be
\la
\delta\phi_1 (\vec x_1, \alpha_1)
\delta\phi_1 (\vec x_3, \alpha')
\ra
=
{1 \over 2\pi^2}
\int_{q_*}^{q_{min}(q_1,q' )} dq q^2 P(q)
{X(q_1)X(q')\over X(q)X(q)}
j_0 (q \vert \vec x_1 - \vec x_3\vert)
\ee
where we defined
$P(q) \equiv {1\over 2} q^{-3} H^2(q)$ where $\alpha = \alpha (q)$.
Using this we can rewrite Eq.(\ref{eq92}) as
\bea
\lefteqn{
{X(q_1)X(q_2)X(q_3)\over 2 m_P}
\int {d^3q'\over (2\pi)^3}
\int {d^3q\over (2\pi)^3}
{P(q') P(q)\over X^2(q)}
	}
\nonumber \\
&  &
\times
W(q'; q_{min}(q_3,q_2))
W(q ; q_{min}(q_1,q' ))
e^{i \vec q \cdot (\vec x_1  - \vec x_3 )}
e^{i {\vec q}' \cdot (\vec x_2  - \vec x_3 )}
\eea
where we defined the filter function $W(q;q_i) \equiv\Theta
(q-q_*)-\Theta (q-q_i)$ ($\Theta$ is the Heaviside function).

A similar analysis can be performed for the term proportional to
$B(\alpha)$ in Eq.(\ref{eqale24}). We get
\bea
\lefteqn{
X(q_1)X(q_2)X(q_3)
\int_{q_*}^{q_3} {dq'\over q'} B(q') X(q')
\int {d^3q''\over (2\pi)^3}
\int {d^3q'''\over (2\pi)^3}
{P(q'')\over X^2(q'')} {P(q''')\over X^2(q''')}
	}
\nonumber \\
&  &
\times
W(q'''; q_{min}(q_2,q'))
W(q'' ; q_{min}(q_1,q'))
e^{i {\vec q}'' \cdot (\vec x_1  - \vec x_3 )}
e^{i {\vec q}''' \cdot (\vec x_2  - \vec x_3 )} \ .
\eea
So far we have been working in configuration space. In order to obtain
the gravitational potential at Hubble radius crossing during the
Friedmann era, it is convenient to Fourier transform the
coarse--grained inflaton fluctuation. One has
$\delta\phi (\vec x, \alpha(q)) = (2\pi)^{-3}
\int d^3k~ \delta\phi (\vec k) \Theta (q - k)
e^{i \vec k \cdot \vec x }$,
where $\delta\phi (\vec k)$ denotes the Fourier transform of the
full scalar field at the time $\alpha(q)$.
This follows from the fact that at the time $\alpha(q)$
the only modes $k$ that contribute to the coarse--grained variable
are those which have already left the inflationary horizon, namely
$k < q$. We can then obtain the Fourier transform
$\delta\phi(\vec k, \alpha(q)) = \delta\phi (\vec k) \Theta (q - k)$,
and, in the limit $k \to q^-$ (or equivalently $q \to k^+$, that is, when
we consider the horizon--crossing time of the given scale),
we simply have $\delta\phi(\vec k, \alpha(k)) = \delta\phi(\vec k)$.
In other words, at the time $\alpha(q)$ the Fourier transform of the
coarse--grained variable coincides with that of the full field.
Using these results in Eq.(\ref{eqale24}) we finally obtain the
inflaton bispectrum through
\bea
\label{field3}
\lefteqn{\la
\delta\phi (\vec k_1, \alpha(k_1))
\delta\phi (\vec k_2, \alpha(k_2))
\delta\phi (\vec k_3, \alpha(k_3)) \ra = (2\pi)^3 \delta^3
(\vec k_1 + \vec k_2 + \vec k_3) P(k_2) P(k_3)
}
\nonumber \\
&  &
\times {X(k_1)\over X(k_2)X(k_3)}
\left[
{ X^2(k_2)\Theta(k_2-k_3) + X^2(k_3)\Theta(k_3-k_2) \over 2 m_P}
+ 2 \int_{k_*}^{k_1} {dq'\over q'} B(q') X(q')
\right]
\nonumber \\
&  &
+ \{\vec k_1\!\leftrightarrow\!\vec k_2\}
+ \{\vec k_1\!\leftrightarrow\!\vec k_3\} \ .
\eea

\subsection{The three--point function of the gravitational potential}

We want now to compute the three--point function of the peculiar
gravitational potential, $\la\Phi(r_0\go)\Phi(r_0\gt)\Phi(r_0\gth) \ra$.
The approximate constancy of the gauge--invariant quantity
$\zeta$ outside the horizon (e.g. Bardeen, Steinhardt \& Turner 1983) allows to
obtain the gravitational potential during the matter--dominated
era, given the value of $\delta\phi$ during inflation.
During inflation one has $\zeta(\vec x,\alpha) \simeq -
\delta\phi(\vec x,\alpha)/(\partial\phi/\partial\alpha)$, which is usually
interpreted as a
linear relation between $\zeta$ and $\delta\phi$. However, when calculating the
three--point function of $\Phi$ one cannot disregard the second--order effects
coming from the fluctuations of $\partial\phi/\partial\alpha$.

Recalling that
\be
{\partial\phi\over\partial\alpha} =
-{m_P^2\over 8\pi}{V'(\phi)\over V(\phi)}\simeq
-{m_P\over 8\pi} \left[ X(\alpha) + X'(\alpha) \delta\phi \right]
\ee
one gets
\be
\label{grapot}
\zeta(\vec x)= {8\pi\over m_P X(\alpha)}
\left(\delta\phi(\vec x ,\alpha)
- {X'(\alpha)\over X(\alpha)} \delta\phi^2 (\vec x,\alpha)
\right) \ .
\ee
This equation is expressed in configuration space;
the Fourier transform of the first term inside the brackets is just
$\delta\phi(\vec k,\alpha)$. For the second term we get a
convolution of the type $\int d^3p \delta\phi(\vec p)
\delta\phi(\vec k - \vec p)$.
For the scales of interest one has $\Phi(\vec k) \simeq -
3\zeta(\vec k)/5$.
Then, adding the two above contributions and evaluating the expression
at the horizon crossing time we get
\begin{eqnarray}
\label{grapot2}
\Phi(\vec k)\!=\! {24\pi \over 5 m_P X(\alpha(k))}
\!\!\left[\!
\delta\phi(\vec k , \alpha(k)\!)\!
- {X'(\!\alpha(k)\!)\over X(\!\alpha(k)\!)}
\!\!\int\!\!\! {d^3p\over (2\pi)^3}
\delta\phi(\vec p , \alpha(k))
\delta\phi(\vec k\! -\! \vec p , \alpha(k))
\!\right].
\end{eqnarray}
{}From this equation we calculate
\bea
\label{bispe}
\lefteqn{ \la \Phi(\vec k_1)\Phi(\vec k_2)\Phi(\vec k_3) \ra =
{ \left(24\pi / 5 m_P \right)^3 \over X(k_1)X(k_2)X(k_3) }
\la \delta\phi(\vec k_1)\delta\phi(\vec k_2)\delta\phi(\vec k_3) \ra +
	}
\nonumber \\
&  &\!
+
{ \left(-24\pi / 5 m_P \right)^3 X'(k_1)
\over
X^2(k_1)X(k_2)X(k_3)
}
\int {d^3p\over (2\pi)^3}
\la
\delta\phi(\vec p )
\delta\phi(\vec k_1 - \vec p)
\delta\phi(\vec k_2 )
\delta\phi(\vec k_3 )
\ra
\nonumber \\
&  &\!
+ \{\vec k_1\!\leftrightarrow\!\vec k_2\}
+ \{\vec k_1\!\leftrightarrow\!\vec k_3\}
\eea
Using the fact that, at horizon crossing, $\la\delta\phi(\vec k_1)
\delta\phi(\vec k_2 ) \ra = (2\pi)^3 P(k_1) \delta^3(\vec k_1 + \vec k_2)$,
we can write
\be
\label{new}
\la \Phi(\vec k_1)\Phi(\vec k_2) \ra = (2 \pi)^3 \delta^3(\vec k_1 +
\vec k_2) f^2(\alpha(k_1)) P(k_1) \ ,
\ee
with $f(\alpha(k)) \equiv 24 \pi/5 m_P X(\alpha(k))$.
Finally, using Eqs.(\ref{new}) and (\ref{field3}) we find
\bea
\label{bispe3}
\lefteqn{
\la \Phi(\vec k_1)\Phi(\vec k_2)\Phi(\vec k_3) \ra =
\left({24\pi\over 5 m_P}\right)^3
(2\pi)^3 \delta^3(\vec k_1 + \vec k_2 + \vec k_3)
{P(k_2)\over X^2(k_2)}
{P(k_3)\over X^2(k_3)}
}
\nonumber \\
&  &\!
\times
\left\{
{X^2(k_2)\Theta(k_2-k_3) + X^2(k_3)\Theta(k_3-k_2) \over 2 m_P}
-
{2X'(k_1)X(k_2)X(k_3)\over X^2(k_1)}
+
\right.
\nonumber \\
&  &
\left.
+
2\int_{k_*}^{k_1} {dq'\over q'} B(q') X(q')
\right\}
 + \{\vec k_1\!\leftrightarrow\!\vec k_2\}
 + \{\vec k_1\!\leftrightarrow\!\vec k_3\}
\eea
We are interested in considering perturbation modes that left the horizon
about 60 e--foldings before the end of the inflationary epoch. In the explicit
examples below $X(k)$ turns out to be a slowly varying function of $k$. We
therefore approximate $X(k) \sim X_{60}$ in what follows. Then, to the lowest
non--vanishing order, the three--point correlation function is
\bea
\label{eqqq}
\lefteqn{\la \Phi (\vec k_1) \Phi (\vec k_2) \Phi (\vec k_3) \ra =
{1 \over f_{60}}\left[
{X_{60} \over 2 m_P}
-
{2 X'_{60} \over X_{60}}
+
{2\over X_{60}}\int_{k_*}^{k_{60}} {dq\over q}B(\alpha(q)) X(\alpha(q))
\right]
	}
\nonumber \\
&  &
\times ~
(2\pi)^3
\delta^3 (\vec k_1 + \vec k_2 + \vec k_3)
\left[
P_{\Phi}(k_1) P_{\Phi}(k_2) + P_{\Phi}(k_2) P_{\Phi}(k_3)
+ P_{\Phi}(k_3) P_{\Phi}(k_1)
\right]
\eea
with $f_{60}\equiv 24\pi / 5 m_P X_{60}$.
Using the expression for $P(q)$ we obtain the power--spectrum for the
peculiar gravitational potential
$P_{\Phi}(k) = 2\pi^2 f_{60}^2 k^{-3} H^2(\alpha (k))/ 4 \pi^2 \simeq
{1 \over 2} f_{60}^2 H_{60}^2 k^{-3} (k/k_*)^{n-1}$,
up to possible presence of small logarithmic
corrections. The primordial spectral index $n$ is related to the
inflationary parameters by the approximate relation
$n \simeq 1 - (X^2_{60}/8\pi) + (m_P X'_{60}/ 4\pi)$
(Turner 1993).
Now we can obtain the two-- and three--point correlation functions for $\Phi$
in configuration space by inverse Fourier transforming Eq.(\ref{new}) and
Eq.(\ref{eqqq}). We then get Eq.(\ref{newer}) and
Eq.(\ref{eq717}) with
\be
{\cal Q}^2 = {8 \pi^2 H_{60}^2 \over 5 m_P^2 X_{60}^2}
{\Gamma (3-n) \Gamma\left(\frac{3}{2}+\frac{n}{2}\right)
\over \left[\Gamma\left(2-\frac{n}{2}\right)\right]^2
\Gamma (\frac{9}{2}-\frac{n}{2})}
\ee
and
\be
\Phi_3 =
{1 \over f_{60}}
\left[
{X_{60} \over 2 m_P}
-
{2 X'_{60} \over X_{60}}
+
{2\over X_{60}} \int_{k_*}^{k_{60}} {dq\over q}
B(\alpha(q)) X(\alpha(q))
\right]
\ee
In order to obtain Eq.(\ref{newer}) and Eq.(\ref{eq717}), we
had to use the definition of ${\cal C}_\ell$. The one given in Section 2
differs from the results obtained here because of the window function
$W(k)$, appearing in the inflationary expressions; this can be however
neglected if we account for the oscillating behaviour of $j_\ell$ for large
arguments and for the fact that $j_\ell\to 0$ for small arguments (that helps
in cancelling the lower divergence).

\section{The CMB skewness}

In what follows we will restrict our analysis to the mean
skewness as given by Eq.(\ref{eq57}), for $\alpha=0$.
In the numerical calculation below we will assume that the higher
multipoles are weighted by a $7^\circ\llap.5$ FWHM beam,
resulting in $\sigma=3^\circ\llap.2$ (e.g. Wright et al. 1992).
To normalize our predictions we will consider the {\em rms} quadrupole
obtained by Seljak \& Bertschinger (1993) through a Maximum--Likelihood
analysis, namely $Q_{rms-PS} = (15.7\pm 2.6) \exp[0.46 (1-n)]\mu K$
(see also Scaramella \& Vittorio 1993 and
Smoot et al. 1993).\footnote{Note that this value
of $Q_{rms-PS}$ assumes a multivariate Gaussian distribution function,
accounting for both the signal and the noise; while in principle
one should repeat the analysis consistently with the assumed statistics of
the temperature perturbations, the quasi--Gaussian nature of our
fluctuation field allows to extrapolate this Maximum--Likelihood
estimate without sensible corrections.}
For the models with $n\lsim 1$, considered below, this expression should be
multiplied by an extra factor $[(3-n)/(14-12n)]^{1/2}$ to
account for the effective decrease in the estimated value of $Q_{rms-PS}$
due to the contribution of gravitational waves (Lucchin, Matarrese \& Mollerach
1992). For the mean temperature we take the FIRAS determination
$T_0=2.726\pm 0.01 K$ (Mather et al. 1993).

To estimate the amplitude of the non--Gaussian character of the fluctuations
we will consider the ``dimensionless" skewness
${\cal S}_1 \equiv \la C_3(0) \ra / \la C_2 (0)\ra ^{3/2}$.
Alternatively, if we want our results to be independent of the normalization,
we may also define the ratio ${\cal S}_2 \equiv \la C_3(0)\ra/\la C_2
(0)\ra^2$, as suggested by the hierarchical aspect of our expression
for the skewness.
We obtain
\be
\label{eq63bis}
{\cal S}_1 =
{ \sqrt{45\pi}\over 32 \pi^2} {\cal Q} X_{60}^2
\left[1-4 m_P {X'_{60}\over X^2_{60}} + {\cal G}\right] {\cal I}_{3/2}(n)
\nonumber
\ee
and
\be
{\cal S}_2 =
\label{eq63}
{15 \over 16 \pi } X_{60}^2
\left[ 1-4 m_P {X'_{60}\over X^2_{60}} + {\cal G} \right] {\cal I}_2(n)
\ee
with
${\cal G} = 4 m_P X_{60}^{-2}
\int_{k_*}^{k_{60}} (dq/q) B(\alpha(q)) X(\alpha(q))$
and where we also defined the spectral index--dependent geometrical factor
\be
\label{eq64}
{\cal I}_p(n)\! =\! {{1\over 3}
\sum_{\ell_1,\ell_2,\ell_3}(2\ell_1\!+\!1)(2\ell_2\!+\!1)(2\ell_3\!+\!1)
\bigl[{\cal C}_{\ell_1} {\cal C}_{\ell_2}\! +\!
{\cal C}_{\ell_2} {\cal C}_{\ell_3}\! +\!
{\cal C}_{\ell_3} {\cal C}_{\ell_1}\bigr]
{\cal W}_{\ell_1}\! {\cal W}_{\ell_2}\! {\cal W}_{\ell_3}
{\cal F}_{\ell_1,\ell_2,\ell_3}
\over \left[\sum_\ell (2l+1) {\cal C}_\ell {\cal W}_\ell^2 \right]^p}
\ee
where the exponent $p$ in the denominator takes values
$3/2$ and $2$ for ${\cal S}_1$ and ${\cal S}_2$, respectively.
The numerical factors ${\cal I}_p(n)$ in Eq.(\ref{eq64}) are
plotted in Figure 3 for different values of the primordial index.

\subsection{Inflationary models}

Let us now specialize our general expressions to some simple inflationary
models.

\begin{flushleft}
{\bf Exponential potential}
\end{flushleft}

Let us first consider power--law inflation driven by the exponential
potential $V(\phi) = V_0 \exp\left(-\lambda\kappa\phi\right)$, with
$\lambda < \sqrt{2}$ (Lucchin \& Matarrese 1985).
In this case  the power--spectrum is an exact power--law
with $n = 1 - 2\lambda^2 / (2 - \lambda^2)$.
We note in passing that the right spectral dependence of the
perturbations can be recovered using the above stochastic approach
(Mollerach et al. 1991). For this model we find $X = - \sqrt{8\pi}\lambda $,
whose constant value implies $A = B = 0$.
We then have
\be
{\cal S}_1
=
{3 \lambda\over 4} ~ {H_{60}\over m_P}~{\cal I}_{3/2}(n)~
\left[
{\Gamma (3-n) \Gamma\left(\frac{3}{2}+\frac{n}{2}\right)
\over \left[\Gamma\left(2-\frac{n}{2}\right)\right]^2
\Gamma (\frac{9}{2}-\frac{n}{2})}
\right]^{1/2}
{}~;~ \ \ \ \
{\cal S}_2 = {15 \over 2 } ~\lambda^2
{}~{\cal I}_{2}(n) \ .
\ee
The COBE results constrain the amplitude of $H_{60}$. For the case $n =
0.8$ we have $H_{60}/m_P = 1.8 \times 10^{-5}$. This gives
${\cal S}_1 =  9.7 \times 10^{-6}$ and
${\cal S}_1 =  1.1\times 10^{-5}$, without and with the
quadrupole contribution respectively, while ${\cal S}_2 =  1.3$ in both
cases.

\begin{flushleft}
{\bf Quartic potential}
\end{flushleft}

Consider now the potential ${1\over 4}\lambda \phi^4$ for chaotic
inflation (Linde 1983, 1985).
For this model $X = 4 m_P/\phi$ and therefore
$m_P B \sim m_P^2 X'' \ll X$. Thus we can take ${\cal G}\sim 0$
in the coefficients of Eqs.(\ref{eq63}) and (\ref{eq63bis})
for the dimensionless skewness parameters.
By using the slow--roll solution for the inflaton we may write
$X \simeq
4 m_P \phi_{60}^{-1}\left[1+\left(2\sqrt{\lambda\over 3}/\kappa
H_{60}\right) \ln(k/k_{60})\right]$ where the logarithmic correction to
the scale invariant power--spectrum is small.
Inflation ends when the inflaton takes the value
$\phi_{end} \simeq \sqrt{2/3\pi} ~m_P$ implying
$\phi_{60} \simeq \sqrt{60/\pi} ~m_P$.
In this case the spectral index is $n \simeq 1$.
We find
\be
{\cal S}_1
=
\sqrt{2 \lambda \over \pi}~{\phi_{60}\over m_P}
{}~{\cal I}_{3/2}(1)
{}~;~ \ \ \ \ \
{\cal S}_2
=
{30 \over \pi }\left({m_P\over \phi_{60}}\right)^2
{}~{\cal I}_{2}(1) \ .
\ee
{\em COBE} constrains the value of $\lambda$ to be $\lambda\simeq 1.4 \times
10^{-13}$. We get ${\cal S}_1 = 5.2\times 10^{-6}$ and
${\cal S}_1 =  6.0\times 10^{-6}$ without and with the quadrupole contribution
respectively, while ${\cal S}_2 = 0.5$.

\begin{flushleft}
{\bf Quadratic potential}
\end{flushleft}

Another simple potential for chaotic inflation is
$\frac12 m_{\phi}^2 \phi^2$ (Linde 1983, 1985). In this case $X \simeq
2 m_P \phi_{60}^{-1}\left[1+\left(2/\kappa^2 \phi_{60}^2\right)
\ln(k/k_{60})\right]$,
$\phi_{60} \simeq \sqrt{30/\pi} ~m_P$
and $n \simeq 1$. We find
\be
{\cal S}_1
=
{3\over 2\sqrt{\pi}}
{}~{m_{\phi}\over m_P}
{}~{\cal I}_{3/2}(1)
{}~;~ \ \ \ \ \
{\cal S}_2
=
{45 \over 4\pi } \left({m_P\over \phi_{60}}\right)^2
{}~{\cal I}_{2}(1)  \ ,
\ee
with $m_{\phi}/m_P \simeq 1.1 \times 10^{-6}$ from the {\em COBE}
normalization. We get ${\cal S}_1 = 3.9 \times 10^{-6} $ and
${\cal S}_1 = 4.5\times  10^{-6} $ without and with the quadrupole
contribution respectively, while ${\cal S}_2 = 0.4$.

\begin{flushleft}
{\bf Hybrid inflation model}
\end{flushleft}

Finally, let us consider a model of hybrid inflation recently
proposed by Linde (1993). Inflation happens in this model during the
slow--roll evolution of the inflaton in the effective
potential $V(\phi)= V_0 + {1\over 2}m^2\phi^2$, where $V_0 = M^4 / 4$
is a cosmological--constant--like term.
At a given time, when the inflaton field takes the value $\phi=M$, its coupling
with a second scalar field triggers a second--order phase
transition of the latter (whose vacuum energy density is responsible for
the cosmological constant term), which makes inflation end.
An interesting prediction of this model is that the spectral index of
density fluctuations for wavelengths which left the
horizon while $V_0>m^2 \phi^2/2$ is larger than unity, namely
$n \simeq 1 + 2 m^2 /\kappa^2 V_0$ (Mollerach et al. 1993).
For this potential we find $X = m_P m^2 \phi / V_0$, implying ${\cal G} = 0$.
We then obtain
\be
{\cal S}_1
=
{\sqrt{3}\over 2}
{\phi_{60}\over m_P}
{m^2\over M^2}
\!\left( 1 - {M^4\over m^2 \phi^2_{60} } \right)
{\cal I}_{3/2}(n)
\left[
{\Gamma (3-n) \Gamma\left(\frac{3}{2}+\frac{n}{2}\right)
\over \left[\Gamma\left(2-\frac{n}{2}\right)\right]^2
\Gamma (\frac{9}{2}-\frac{n}{2})}
\right]^{1/2}
\ee
\be
{\cal S}_2
=
{15\over\pi }
{m_P^2 m^4 \phi_{60}^2 \over M^8}
\!\left( 1 - {M^4\over m^2 \phi^2_{60} } \right)
{\cal I}_{2}(n)
\nonumber
\ee
The slow--roll solution for the inflaton gives
$\phi_{60} = M \exp [60~ m^2 /\kappa^2 V_0]$. Choosing a value
$n=1.1$, the {\em COBE} results constrain the free parameters to be
$M \simeq 1.3 \times 10^{-4} m_P$ and $m \simeq 10^{-8} m_P$.
In this case we find ${\cal S}_1 = - 1.1\times 10^{-5}$
and ${\cal S}_1 = - 1.3\times 10^{-5}$
without and with the quadrupole contribution, while
${\cal S}_2 = - 1.6$. Note that these results suggest some correlation
between the sign of the skewness and the value of the spectral index
$n$.

\section{Conclusions}

The results reported in the previous section seem to preclude any chance of
actually obtaining observable non--Gaussian signals at least in the frame
of inflation, unless one resorts to more complicated multiple--field
models (e.g. Allen, Grinstein \&
Wise 1987; Kofman et al. 1991; Salopek 1992).
However, it should be stressed that all the results reported in Section 2
would apply to any large--scale anisotropy where the temperature fluctuation
can be obtained by a local perturbative calculation (this is not the case, for
instance, for
secondary anisotropies, which are to be ascribed to an integrated effect):
under these conditions the {\em hierarchical} form of Eq.(\ref{eq717}),
where the bispectrum is a sum of products of two power--spectra, holds.

One may also wonder whether
the general analysis of the CMB skewness generated in the frame of
the inflationary model may provide further constraints on the model
parameters. We did not address this issue in the present paper, where
we only considered some simple realistic models. In these cases
a relevant constraint can be obtained.
Both for the exponential potential and for the polynomial ones
the skewness can be estimated by taking just the first term in
the brackets of Eqs.(\ref{eq63bis}) and (\ref{eq63}).
An upper limit on its magnitude is then obtained by requiring that they
give rise to an accelerated universe expansion (i.e. inflation), which
provides a constraint on the steepness of the potential, $X^2 < 24\pi$. This
corresponds to ${\cal S}_1 \lsim  10^{-4}$ and ${\cal S}_2 \lsim 20$.

Notwithstanding the non--zero value of the skewness, any actual
detection of this signal is hardly distinguishable from
the cosmic variance noise in which it is embedded: the
intrinsic limitation induced by our impossibility of making
measurements in more than one universe.
In fact, the overall coefficient of ${\cal S}_1$ is generically
much smaller than the dimensionless {\em rms} skewness calculated from an
underlying Gaussian density field. Removing the quadrupole contribution
to reduce the cosmic {\em rms} skewness, as recently suggested by Luo (1993b),
does not change this main conclusion, because the predicted mean skewness
would also be reduced by a comparable factor.
Things get even worse if we take into account that no sampling
of the CMB sky is complete, not even the {\em COBE} one, where a cut for
galactic latitudes $\vert b\vert < 20^\circ$ is required.
As shown by Scott et al. (1993), this ``sample variance'' contributes in
hiding the signal. Therefore, any possible skewness detection on the
{\em COBE} scale is most probably due to these statistical effects than to any
primordial non--Gaussian feature.

\vspace{5mm}

\noindent{\bf Acknowledgement:} The work of A. Gangui was supported in
part by the Italian MAE. F. Lucchin and S. Matarrese acknowledge
financial support from the Italian MURST.
The work of S. Mollerach was supported by a grant from the Commission of
European Communities (Human capital and mobility programme).

\vspace{1cm}

\section*{References}

\vspace{.2in}

\begin{description}

\item[\rm Abbott, L.F. \& Wise, M.B.] 1984, ApJ, 282, L47

\item[\rm Allen, T.J., Grinstein, B. \& Wise, M.B.] 1987, Phys. Lett. B, 197,
66

\item[\rm Bardeen, J.M., Steinhardt, P.J. \& Turner, M.S.] 1983,
Phys. Rev., D28, 679

\item[\rm Barrow, J.D. \& Coles, P.] 1990, MNRAS, 244, 188

\item[\rm Bond, J.R. \& Efstathiou, G.] 1987, MNRAS, 226, 655

\item[\rm Coulson, D., Ferreira, P., Graham, P. \& Turok, N.] 1993, preprint
(PUP--TH--1429)

\item[\rm Coulson, D., Pen, U.--L. \& Turok, N.] 1993, preprint (PUP--TH--1393)

\item[\rm Fabbri, R., Lucchin, F. \& Matarrese, S.] 1987, ApJ, 315, 1

\item[\rm Falk, T., Rangarajan, R. \& Srednicki, M.] 1993, ApJ, 403, L1

\item[\rm Goncharov, A.S., Linde, A.D. \& Mukhanov, V.F.] 1987, Int. J. Mod.
Phys., A2, 561

\item[\rm Graham, P., Turok, N., Lubin, P.M. \& Schuster, J.A.] 1993, preprint
(PUP--TH--1408)

\item[\rm Hinshaw, G. et al.] 1993, ApJ, submitted (preprint astro--ph/9311030)

\item[\rm Kofman, L.K., Blumenthal, G.R., Hodges, H. \& Primack, J.R.]
1990, in Proc. Workshop on {\em Large--Scale Structures and Peculiar Motions in
the Universe}, Rio de Janeiro, May 1989, Latham, D.W. \& da Costa, L.N. eds.,
ASP Conference Series, Vol. 15

\item[\rm Linde, A.] 1983, Phys. Lett. B, 129, 177

\item[\rm Linde, A.] 1985, Phys. Lett. B, 162, 281

\item[\rm Linde, A.] 1993, preprint (astro--ph/9307002)

\item[\rm Lucchin, F. \& Matarrese, S.] 1985, Phys. Rev., D32, 1316

\item[\rm Lucchin, F., Matarrese, S. \& Mollerach, S.] 1992, ApJ, 401, L49

\item[\rm Luo, X.] 1993a, Phys. Rev. D, submitted (FNAL--PUB--93/294--A)

\item[\rm Luo, X.] 1993b, ApJ, submitted (preprint astro--ph/9312004)

\item[\rm Luo, X. \& Schramm, D.N.] 1993, Phys. Rev. Lett., 71, 1124

\item[\rm Mart{\'\i}nez--Gonz\'alez, E., Sanz, J.L. \& Silk, J.] 1992,
Phys. Rev. D46, 4193

\item[\rm Mather, J. et al.] 1993, ApJ, in press

\item[\rm Messiah, A.] 1976, {\em Quantum Mechanics}, Vol.2
(Amsterdam: North--Holland)

\item[\rm Moessner, R., Perivolaropoulos, L. \& Brandenberger, R.] 1993
preprint (BROWN--HET--911)

\item[\rm Mollerach, S., Matarrese, S. \& Lucchin, F.] 1993, ApJ, submitted
(preprint astro--ph/9309054)

\item[\rm Mollerach, S., Matarrese, S., Ortolan, A. \& Lucchin, F.] 1991,
Phys. Rev. D44, 1670

\item[\rm Rees, M. \& Sciama, D.W.] 1968, Nature, 217, 511

\item[\rm Sachs, R. \& Wolfe, A.] 1967, ApJ, 147, 73

\item[\rm Salopek, D.S.] 1992, Phys. Rev. D45, 1005

\item[\rm Scaramella, R. \& Vittorio, N.] 1991, ApJ, 375, 439

\item[\rm Scaramella, R. \& Vittorio, N.] 1993 MNRAS, 263, L17

\item[\rm Scott, D., Srednicki, M. \& White, M.] 1993, ApJ, submitted
(preprint astro--ph/9305030)

\item[\rm Seljak, U. \& Bertschinger, E.] 1993, ApJ, 417, L9

\item[\rm Smoot, G.F. et al.] 1992, ApJ, 396, L1

\item[\rm Smoot, G.F. et al.] 1993, ApJ, submitted (preprint astro--ph/9312031)

\item[\rm Srednicki, M.] 1993, ApJ, 416, L1

\item[\rm Starobinskii, A.] 1986, in {\em Field Theory, Quantum Gravity
and Strings}, Proceedings of the Seminar Series, Meudon and Paris, France,
1984--1985, de Vega, H.J. \& Sanchez, N. eds., Lecture Notes in Physics
Vol. 246 (Berlin: Springer--Verlag)

\item[\rm Turner, M.S.] 1993, Phys. Rev. D48, 3502

\item[\rm Wright, E.L. et al.] 1992, ApJ, 396, L13

\item[\rm Yi, I. \& Vishniac, E.T.] 1993, Phys. Rev. D48, 950

\end{description}

\vspace{.2in}

\newpage

\section*{Figure captions}

\begin{description}

\item[Figure 1:] Collapsed three--point function (with and without the
quadrupole contribution), normalized to the skewness, as a function of the
angular separation $\alpha$ (left panel). The different curves refer to three
typical values of the primordial spectral index: $n=0.8,~1,~1.2$.
The symbols in the left panel are the same as in the right one, which shows a
vertical expansion of the same plot.

\item[Figure 2:] Normalized {\it rms} skewness of a Gaussian temperature
fluctuation field as a function of the spectral index $n$, both including
and removing the quadrupole contribution.

\item[Figure 3:] Geometrical factors ${\cal I}_p(n)$ as a function of $n$.
Crosses are for ${\cal I}_{3/2}(n)$, where the quadrupole term is removed.
Diamonds correspond to ${\cal I}_{3/2}(n)$ including the quadrupole
contribution. Open boxes correspond to ${\cal I}_{2}(n)$. For the latter
case the difference in including or removing the quadrupole contribution
is ${\cal O}(10^{-2})$, which is not perceivable in this graph.

\end{description}

\end{document}